\documentclass[twocolumn]{article}
\usepackage[a4paper,left=1.5cm,right=1.5cm,top=3cm,bottom=4cm]{geometry}

\usepackage{amsmath,amssymb,amsfonts}   
\usepackage[bottom]{footmisc}

\usepackage{algorithmic}
\usepackage{hyperref}
\usepackage{graphicx}
\usepackage{subcaption}
\usepackage[sort&compress,super,square,comma]{natbib}
\usepackage{tabularx}
\usepackage{booktabs}
\usepackage{multirow}
\usepackage{float}
\usepackage{textcomp}
\usepackage{xcolor}
\usepackage[version=4]{mhchem}
\usepackage{enumerate}
\usepackage{enumitem}
\usepackage{siunitx}
\usepackage{upgreek}
\usepackage{authblk}

\usepackage{amsmath,amssymb}

\makeatletter
\newcommand{\@defs@vec}[1]{\boldsymbol{#1}}
\newcommand{\@defs@tens}[1]{\mathbf{#1}}

\newcommand{\fnorm}[1]{| #1 |\@defs@replaced{abs}} 

\newcommand{\vect}[1]{\@defs@vec{#1}{}}       
\newcommand{\tens}[1]{\@defs@tens{#1}{}}      

\newcommand{\nvec}{\@defs@vec{n}}

\newcommand{\rf}{\text{ref}}      

\newcommand{\SEI}{\text{SEI}}

\newcommand{\diff}{{\text{diff}}}
\newcommand{\Lint}{{\text{e}^{\text{-}}}}

\newcommand{\kB}{k_\text{B}}

\newcommand{\cref}{c_\text{max}}

\makeatother

\makeatletter
\renewcommand*{\@fnsymbol}[1]{\ensuremath{\ifcase#1  \or = \or * \or \dagger\or \ddagger\or 
   \mathsection\or \mathparagraph\or * \or \|\or **\or \dagger\dagger
   \or \ddagger\ddagger \else\@ctrerr\fi}}
\makeatother

\begin{document}

\title{Transition between growth of dense and porous films: Theory of dual-layer SEI}

\date{\today}

\author[1,2]{Lars~von~Kolzenberg \thanks{These authors contributed equally to this work}}
\author[1,2]{Martin~Werres $^=$}
\author[3]{Jonas Tetzloff}
\author[1,2,3]{Birger~Horstmann \thanks{Corresponding Author: birger.horstmann@dlr.de}}
\affil[1]{German Aerospace Center, Pfaffenwaldring 38-40, 70569 Stuttgart, Germany}
\affil[2]{Helmholtz Institute Ulm, Helmholtzstra{\ss}e 11, 89081 Ulm, Germany}
\affil[3]{Ulm University, Albert-Einstein-Allee 47, 89081 Ulm, Germany}

\maketitle

\begin{abstract}
The formation of passivating films is a common aging phenomenon, for example in weathering of rocks, silicon, and metals. In many cases, a dual-layer structure with a dense inner and a porous outer layer emerges. However, the origin of this dual-layer growth is so far not fully understood. In this work, a continuum model is developed, which describes the morphology evolution of the solid-electrolyte interphase (SEI) in lithium-ion batteries. Transport through the SEI and a growth reaction governed by the SEI surface energies are modelled. In agreement with experiments, this theory predicts that SEI grows initially as a dense film and subsequently as a porous layer. This dynamic phase transition is driven by the slowing down of electron transport as the film thickens.  Thereby, the model offers a universal explanation for the emergence of dual-layer structures in passivating films.
\end{abstract}

\section{Introduction}

Surface processes are at the center of various aging phenomena. Weathering forms a patina on earth \cite{Pineda1990,Garcia-Valles1998} and even moon minerals \cite{Wentworth1999}, as well as on various metals including copper \cite{Payer1995,Robbiola1998,Macdonald1999,Zhang2002,Sandberg2006,Souissi2006,Muresan2007,Hernandez2011,Travassos2019,Domenech-Carbo2017,Domenech-Carbo2019}, zinc \cite{Macdonald1998,Macdonald1999,Thomas2013}, and lead \cite{Macdonald1999,Domenech-Carbo2017,Domenech-Carbo2019}. Crystalline silicon passivates in contact with hydrogen \cite{yoshikawa2017silicon,masuko2014achievement,de2012high,tanaka1992development,van2012introduction,Nunomura2020}, nitride \cite{aberle2001overview,aberle2000surface}, or oxygen \cite{larionova2010surface,kerr2001surface}.
Also anode particles in lithium-ion batteries form a shielding surface film; the solid-electrolyte interphase (SEI) \cite{Peled1979,Peled2017,Wang2018}. Different mechanisms for SEI growth have recently been discussed with continuum models \cite{Horstmann2018,VonKolzenberg2020,Das2019,Harris2020,Hao2017,Krauss2022}.

Despite the variety of substrates and aging conditions, some universal properties of these surface films emerge. First, the film protects the bulk material and thereby slows down its own growth over time \cite{Macdonald1999,Wang2018,Horstmann2018,Domenech-Carbo2019}. This self-passivating characteristic is mostly ascribed to transport limitations through the film \cite{Deal1965,Robbiola1998,Macdonald1999,Domenech-Carbo2014,Domenech-Carbo2019,Horstmann2018,Single2016,Single2017,Single2018,Chazalviel2010,VonKolzenberg2020,Graedel1996,Tidblad1996,Tidblad1996a,Larson2002}.
Second, the film structure is oftentimes dual-layered with a dense inner layer and a porous outer layer \cite{Payer1995,Atrens1996,Macdonald1998,Macdonald1999,Thomas2013,Domenech-Carbo2017,Domenech-Carbo2019,Nunomura2020,Lu2011,Harris2013,Lu2014,Peled2017}.

The origin of this multi-layer growth remains to be understood and only few theoretical works analyze the surface morphology evolution. Kim and Kosterlitz \cite{Kim1989} derived a scaling law to describe a ballistic growth process via a solid-on-solid model.
Clarelli \textit{et al.} \cite{Clarelli2014} modeled copper weathering and showed the thickness evolution of the inner cuprite and the outer brochantite surface film.
Chazalviel \textit{et al.} \cite{Chazalviel2010} derived a threshold voltage for the transition from uniform to mesoporous growth of $\ce{SiO_2}$ in fluorinated media using a diffusion-reaction model.
Single \textit{et al.} \cite{Single2016,Single2017} modeled the SEI porosity evolution with two complementary transport processes and two different formation reactions. 

In this paper, we present a novel model to study the morphology evolution of surface films for the example of SEI growth. The model assumes diffusion as growth limiting process \cite{Shi2013,Soto2015,Single2018}. The subsequent film formation reaction is governed by the film morphology according to the model of Horstmann \textit{et al.} for the formation of $\ce{Li_2O_2}$ films and particles in lithium-air batteries \cite{Horstmann2013,doi:10.1063/5.0006833}. This approach can be understood in more general terms as a model for electrochemically-driven phase transitions due to electro-autocatalysis \cite{Bazant2017}. In the following, we first present the theoretical foundations of our model before presenting and discussing its numeric evaluation.

\section{Theory}

Passivating films grow at their interface with the environment. 
In the case of SEI, new film forms from non-aqueous electrolyte, lithium ions $\ce{Li^+}$, and electrons $\ce{e^-}$.
We simplify the multitude of possible SEI growth reactions to the formation of the most prominent SEI compound lithium ethylene dicarbonate $\ce{Li_2EDC}$ from the solvent ethylene carbonate ($\ce{EC}$),
\begin{equation}
    \label{eq:Reaction_Equation}
    \ce{Li^+}+\ce{e^-}+ \ce{EC} \longrightarrow 0.5\ce{Li_2EDC} + 0.5\ce{C_2H_4}.
\end{equation}

We model the kinetics $r$ of Reaction \ref{eq:Reaction_Equation} with the thermodynamically consistent approach \cite{Latz2013,Bazant2013}
\begin{equation}
    \label{eq:Reaction_Modeling}
    r = r_0  \left(e^{\frac{\mu_\Lint}{\kB T}}-e^{\frac{\mu_\SEI}{2\kB T}} \right),
\end{equation}
with the rate constant $r_0$, the Boltzmann constant $k_\text{B}$, and the temperature $T$ in Kelvin. We assume diffusion of localized electrons, for example \textit{via} lithium interstitial atoms, $\ce{Li^0}=\ce{Li^+}+\ce{e^-}$ \cite{Shi2013,Soto2015}, as rate limiting step. The additional assumption of constant lithium ion and electrolyte concentration at the SEI-electrolyte interface simplifies the educt chemical potential to $\mu_\Lint$ in Equation \ref{eq:Reaction_Modeling}. In the following, we model morphology dependent SEI thermodynamics and electron transport through the SEI. Finally, we perform a stability analysis of SEI growth.

\begin{figure}[tb]
    \centering
    \includegraphics[width=8.4cm]{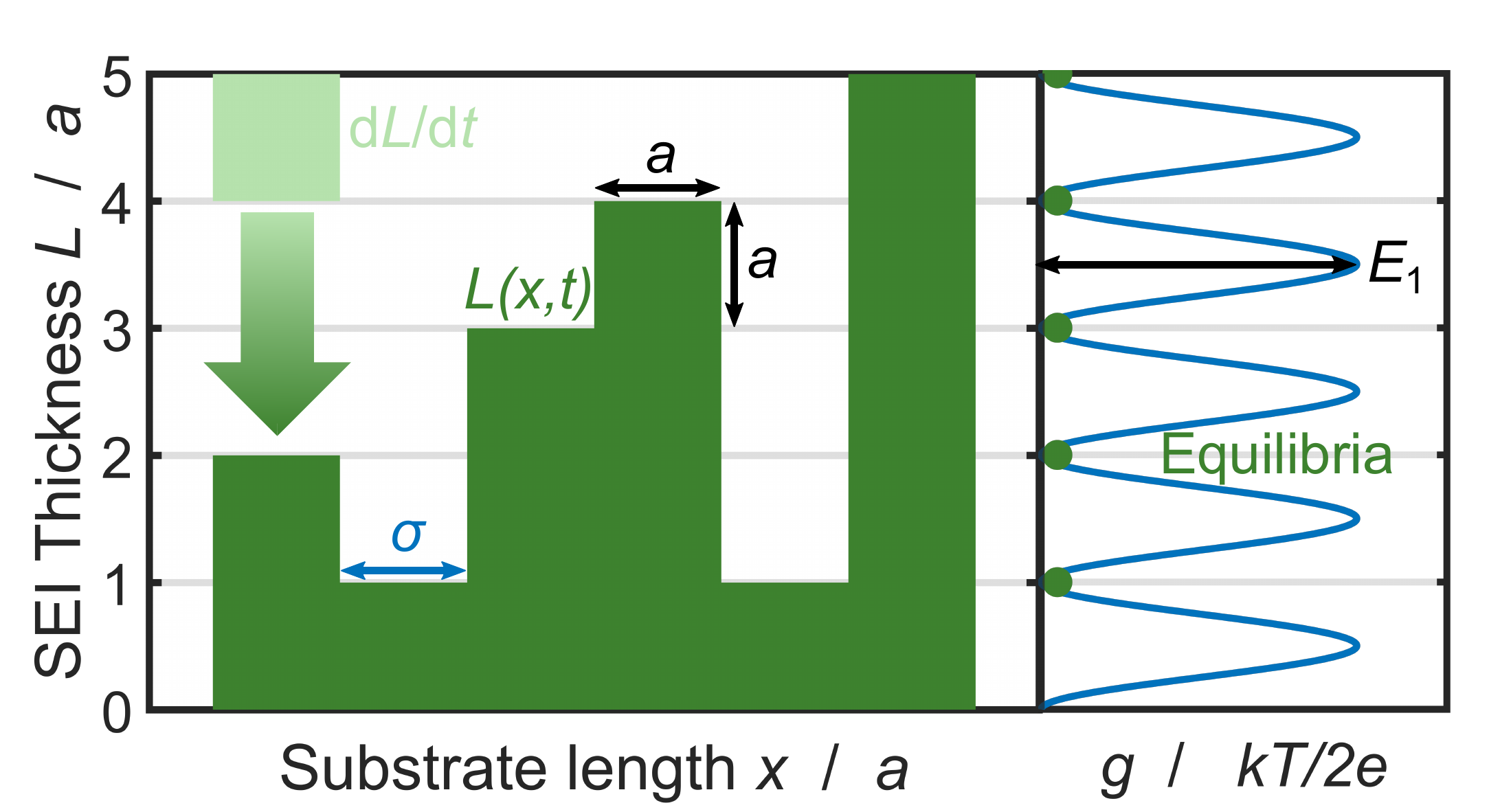}
    \caption{Schematic depiction of the SEI growth process and the corresponding geometrical quantities. Left: SEI molecules in the form of cubes with edge length $a$ deposit in time $\text{d}L/\text{d}t$ (bright green) resulting in the SEI thickness $L(x,t)$ (dark green) governed by Equation \ref{eq:SEI_Growth}. 
    Right: The oscillating bulk contribution of the Gibbs free energy density $g(L)$ for the SEI morphology. Equilibria of the Gibbs free energy density $g$ lie on integer monolayers and non-integer monolayer deposition is penalized with the energy $E_1$ (see Equation \ref{eq:Bulk_Free_Energy}).}
    \label{fig:Morphology_Sketch}
\end{figure}

Now, we discuss how we represent SEI morphology in our model. Figure \ref{fig:Morphology_Sketch} illustrates the SEI growth process and the corresponding geometrical quantities. We model the two-dimensional substrate surface as a one-dimensional coordinate $x$. On a molecular level, the SEI grows \textit{via} the deposition of distinct molecules, which we model as cubes of edge length $a=\SI{5.42}{\angstrom}$ \cite{Borodin2006}. Each electron transferred in reaction \ref{eq:Reaction_Equation} increases the local SEI thickness $L(x)$ by half a monolayer $a/2$ (green area in Figure \ref{fig:Morphology_Sketch}). Thus, the number of monolayers $\tilde{L}=L/a$ grows according to
\begin{equation}
    \label{eq:SEI_Growth}
    \frac{\partial \tilde{L}(x,t)}{\partial t}= \frac{\alpha}{2}  a^2 r,
\end{equation}
with the surface area $a^2$ of each molecular site.
Inhomogeneous deposition curves the SEI surface $s$ (boundary of green area in Figure \ref{fig:Morphology_Sketch}) and thus changes the surface increment depending on the geometric factor $\alpha=\text{d}s/\text{d}x=\sqrt{1+(\partial L/\partial x)^2}$.

The chemical potential of the SEI $\mu_\SEI$ depends on the resulting SEI morphology \textit{via} the variational derivative of the Gibbs free energy $\mu_\SEI=\delta G/\delta \tilde{L}$, see for example Horstmann \textit{et al.} \cite{Horstmann2013}.
The free energy $G$ consists of a surface $g_\text{s}$ and a bulk contribution $g_\text{b}$ according to Equation \ref{eq:Free_Energy},
\begin{equation}
    \label{eq:Free_Energy}
    G = \int g_\text{b} \text{d}x + \int g_\text{s} \text{d}s.
\end{equation}
The surface free energy density $g_\text{s}=\sigma$ accounts for the surface energy $\sigma =\SI{10}{\milli\electronvolt}/a$ and acts along the curved surface $\text{d}s=\alpha\cdot\text{d}x$ of the SEI \cite{Stone2005}. 

The bulk free energy density,
\begin{equation}
    \label{eq:Bulk_Free_Energy}
    g_\text{b}=\frac{2e}{ a } \left(-E_0 \tilde{L} + \frac{E_1}{\pi} \sin^2\left(\pi \tilde{L}\right) \right)
\end{equation}
contains two different terms.
The first term accords to the typical SEI formation voltage $E_0=\SI{0.8}{\volt}$ \cite{Horstmann2013,Luchkin2020}. The
second term reflects the molecule-wise deposition of SEI molecules and as such penalizes thicknesses apart from integer monolayers, see Figure \ref{fig:Morphology_Sketch} right \cite{Horstmann2013}. Because $\ce{Li_2EDC}$ exhibits an amorphous structure, we use the small energy barrier $E_1=\SI{10}{\milli\volt}$. 

We model molecular disorder by adding a normally distributed random term with an amplitude of $\SI{1}{\milli\volt}$ to this barrier. This disorder term accounts for structural variations in SEI molecule deposition and as such varies for each monolayer in thickness $\tilde{L}$ and substrate $\tilde{x}=x/a$ direction \cite{Horstmann2013,Kim1989}. 
This source of randomness breaks the symmetry of the system and thus drives the emergence of instabilities:
Locally higher energy barriers provide less favorable sites for deposition of distinct SEI molecules. In turn, this structural heterogeneity affects the reaction rate in Equation \ref{eq:Reaction_Modeling} and thus either enhances or diminishes unstable growth.

Next, we model electron transport through the SEI \textit{via} diffusion of neutral lithium as proposed in Ref. \citenum{Single2018,VonKolzenberg2020}.
The chemical potential of lithium atoms inside the SEI $\mu_\Lint$ follows from the Nernst equation $\mu_\Lint=\kB T\ln\left(c/c_\rf\right)$, where the reference concentration $c_\rf$ is a material parameter. Lithium atoms diffuse through the SEI at the diffusive flux density
\begin{equation}
    \label{eq:Diffusion}
    N_\diff = -D\cdot\text{grad}~ c_\Lint,
\end{equation}
yielding the concentration $c(x)$ at the SEI surface.

In the following, we simplify the flux equation to one dimension. We emphasize that we want to focus on the onset of porous layer growth in this paper. Thus, we discuss small fluctuations in concentrations $\delta c(x)$ due to small fluctuations in SEI thickness $\delta L$ around a homogeneous film with thickness $L$ in this paragraph. The diffusive transport in Equation \ref{eq:Diffusion} leads to a decrease in concentration fluctuation with mean thickness as $\delta c / \delta L \sim c_0 / L$. We conclude that the diffusive concentration term in Equation \ref{eq:Reaction_Modeling} becomes less and less important with growing SEI. Therefore, we simplify the model equations and simulate transport in one dimension only. We make the simplified Ansatz of a constant surface concentration $\delta c / \delta L=0$, which we determine with $N_\diff = -D(c-c_0)/L$ and the integral condition
\begin{equation}
    \label{eq:Integral_Closing}
    \bar{N}_\diff = \frac{1}{A_x}\int_0^{A_x} N_\diff \text{d}x = \frac{1}{A_x}\int_0^{A_x} r \text{d}s = \bar{r},
\end{equation}
where the integral ranges over the substrate length $A_x$.
The concentration $c_0=c_\rf \exp\left(-eU_0/\kB T \right)$ at the anode-SEI interface depends on the open circuit voltage (OCV) $U_0$, which results from the state of charge SoC \textit{via} the OCV-curve \cite{Single2018,VonKolzenberg2020,VonKolzenberg2021}.

\begin{center}
\begin{table}[tb]
\begin{tabular}{ccc}
\toprule
     $\tilde{L}=L/a$ & $\tilde{x}=x/a$ & $\tilde{A}_x=A_x/a$  \\
     $\tilde{t}=tr_0a^2$  & $\tilde{c}=c/c_\rf$ &  $\tilde{\kappa}=a\sigma/\kB T$ \\
     $\tilde{E}_0=2eE_0/\kB T$ & $\tilde{E}_1=2eE_1/\kB T$ &
    $\tilde{U}_0=eU_0/\kB T$ \\ $\text{Da}_\text{II}=r_0a/Dc_\rf$ \\
\bottomrule
\end{tabular}
\caption{Normalized quantities}
\label{tab:Normalized_Parameters}
\end{table}
\end{center}

At this point, we summarize and non-dimensionalize the model equations. We use the characteristic time $1/(r_0 a^2)$ and the characteristic energy scale $\kB T$.
Table \ref{tab:Normalized_Parameters} lists the dimensionless parameters of our model.
\begin{gather}
    \label{eq:Growth_Summary}
    \frac{\partial \tilde{L}}{\partial \tilde{t}}=\frac{1}{2}\alpha\left(\tilde{c}-e^{\tilde{\mu}_\SEI/2}\right) \\
    \label{eq:Growth_Summary_2}
         \tilde{\mu}_\SEI = -\tilde{E}_0 + \tilde{E}_1\sin(2\pi \tilde{L}) -\frac{\tilde{\kappa}}{\alpha^3}\frac{\partial^2\tilde{L}}{\partial \tilde{x}^2} \\
         \label{eq:Growth_Summary_3}
         \tilde{c}= \frac{\int_0^{\tilde{A}_x} e^{-\tilde{U}_0}/\tilde{L} + \text{Da}_\text{II} \alpha e^{\tilde{\mu}_\SEI/2}\text{d}x}{\int_0^{\tilde{A}_x} 1/\tilde{L} + \text{Da}_\text{II} \alpha \text{d}x}
\end{gather}
We solve the differential Equation \ref{eq:Growth_Summary} using the solver \textit{ode15s} in MATLAB with the parameters listed in the supporting information SI-1 \cite{Borodin2006,Luchkin2020,Birkl2015,Keil2016,Keil2017}. The parameters $c_\text{ref}$, $D$ and $r_0$ are fitted to the storage experiments of Keil \textit{et al.} \cite{Keil2016,Keil2017}, as explained detailedly in supporting information SI-2.

Linear stability analysis mathematically describes the response of a dynamic system to infinitesimal perturbations \cite{Horstmann2013}. We start the analysis with a homogeneous thickness profile $\tilde{L}_0$ so that $\alpha=1$. This profile is perturbed by the infinitesimal function $\delta \tilde{L}(x)$, which satisfies $\partial^2 \delta \tilde{L}/\partial \tilde{x}^2=-\tilde{k}^2\delta\tilde{L}$ with the dimensionless wavenumber $\tilde{k}=ka$.
We evaluate the growth of this infinitesimal perturbation with Equation \ref{eq:Growth_Summary},
\begin{equation}
    \label{eq:Stability_Analysis_1}
    \frac{\partial \delta \tilde{L}}{\partial \tilde{t}}=-\frac{e^{\tilde{\mu}_\SEI/2}}{4} \frac{\delta \tilde{\mu}_\SEI}{\delta \tilde{L}}\delta \tilde{L}=\tilde{s}\cdot\delta \tilde{L}.
\end{equation}
Here, we use that the surface concentration is not perturbed $\delta \tilde{c}/\delta \tilde{L}=0$ as discussed and assumed in the context of Equation \ref{eq:Diffusion} above.

Evaluating this derivative, we obtain the exponential growth rate 
\begin{align}
    \label{eq:Exponential_Growth_Rate}
    \tilde{s} = -\frac{1}{4}e^{\tilde{\mu}_{\SEI,0}/2}\cdot&\left(2\pi\tilde{E}_1\cos(2\pi\tilde{L}_0)+\tilde{k}^2\tilde{\kappa}\right).
\end{align}
SEI growth is unstable if the perturbation grows faster than the homogeneous film, \textit{i.e.}
\begin{equation}
    \label{eq:instability_criterion}
    \tilde{s}> \partial{\tilde{L}}_0/\partial \tilde{t} \ .
\end{equation}
The instability criterion \ref{eq:instability_criterion} describes that growth becomes unstable if local fluctuations in the SEI thickness emerge faster than the mean SEI grows. In this case, the fluctuations dominate the overall growth and lead to a heterogeneous structure. We refer to Ref. \citenum{Horstmann2013} for further discussions.

\section{Results}

In this section, we discuss our simulation results and compare them with experiments. Our simulations predict the capacity fade due to mean SEI growth and the SEI dual-layer structure.

\begin{figure}[tb]
    \centering
    \includegraphics[width=8.4cm]{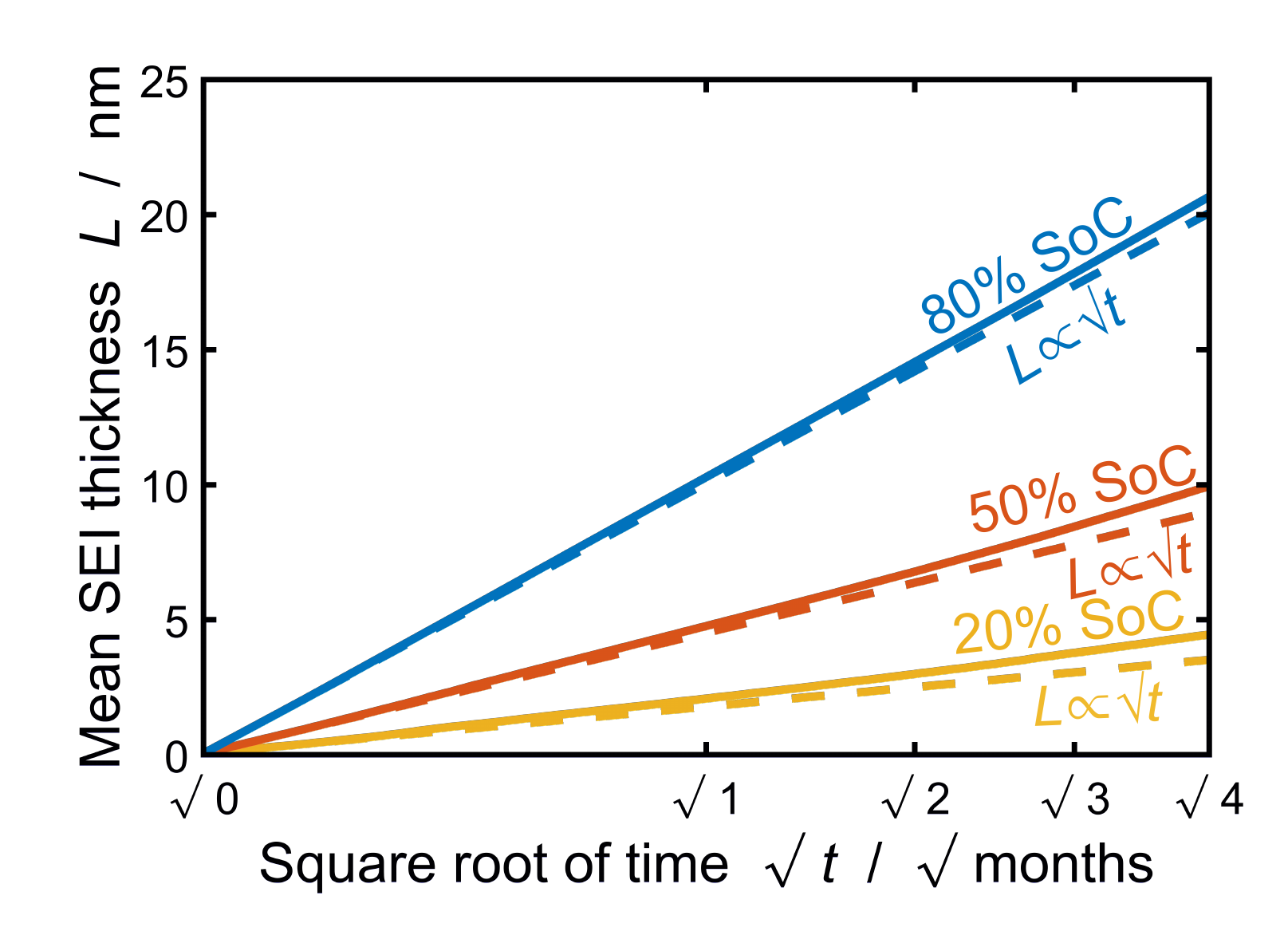}
    \caption{Growth of the mean SEI thickness $\bar{L}_\SEI=1/A_x\int_0^{A_x}L_\SEI\text{d}\tilde{x}$ over the square root of storage time $\sqrt{t}$. Different colors indicate different states of charge. Dashed lines indicate a $\sqrt{t}$-growth for comparison.}
    \label{fig:sqrt_t_growth}
\end{figure}
It is well known in experiment and theory that SEI growth during storage leads to measurable capacity fade. In Figure \ref{fig:sqrt_t_growth}, we plot the mean SEI thickness $\bar{L}=1/\tilde{A}_x\int_0^{\tilde{A}_x}L\text{d}\tilde{x}$ over time. Our model reproduces two experimentally observed trends. First, the SEI grows faster for higher states of charge SoC. This is in line with the experiments of Keil \textit{et al.} \cite{Keil2016,Keil2017}, which we further validate in the supporting information SI-2. 
Second, the SEI grows proportional to the square root of time $\sqrt{t}$, which accords well to battery aging studies \cite{Broussely2001}. However, for low SoCs and large storage times, we observe a deviation from the $\sqrt{t}$-growth. 

\begin{figure}[tb]
    \centering
    \includegraphics[width=8.4cm]{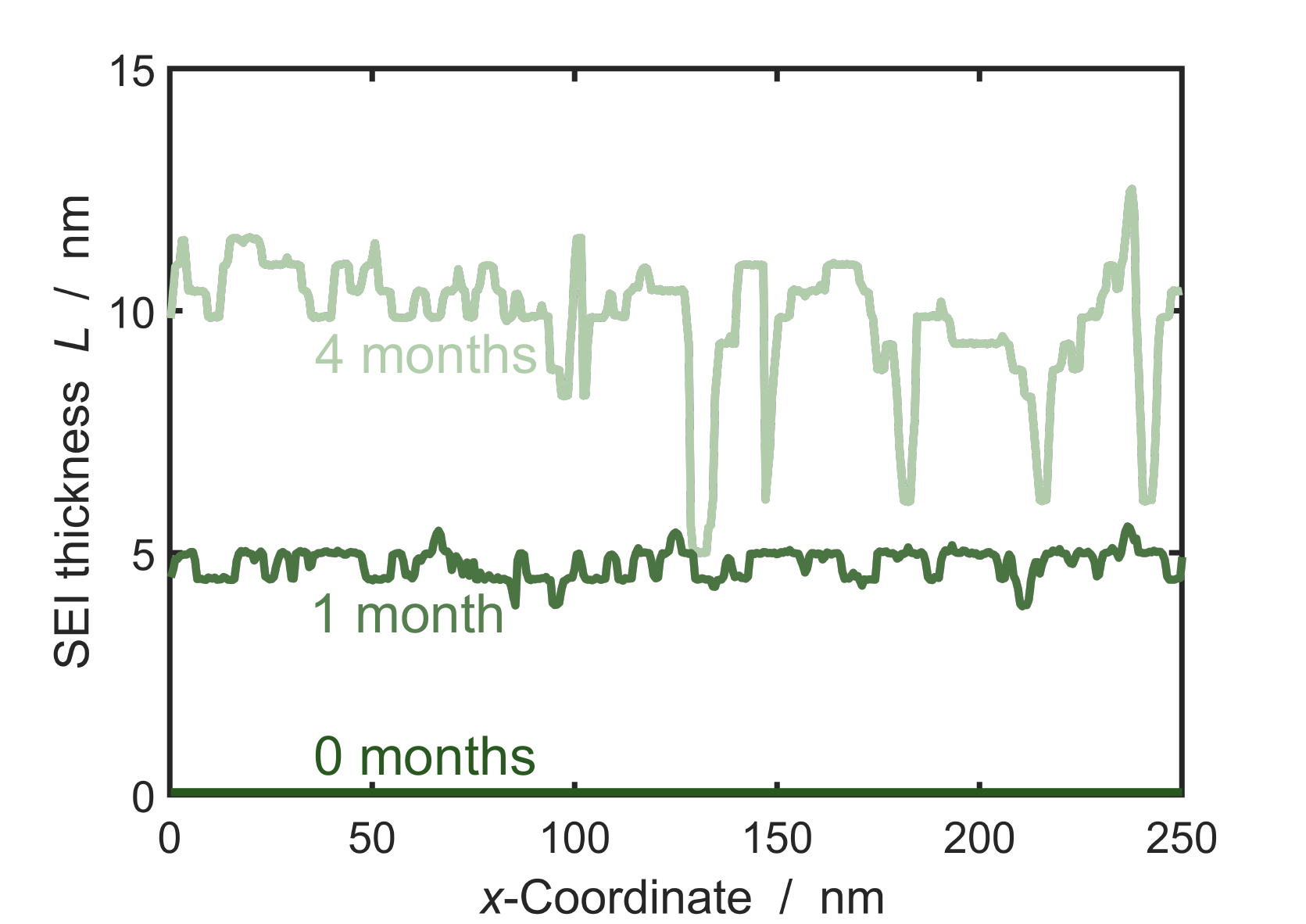}
    \caption{SEI morphology evolution during storage at 50\% state of charge described by the SEI thickness $L(x)$ over the substrate $x$. Color gradient indicates increasing storage time from 0 (darkest green) to 4 months (lightest green).}
    \label{fig:Morphology_Evolution}
\end{figure}
In the following, we study SEI morphology evolution. Figure \ref{fig:Morphology_Evolution} shows SEI morphologies at SoC=50\% for a couple of storage times. The SEI grows from an initially homogeneous profile at 0 months to an increasingly heterogeneous profile after 4 months. We interpret the heterogeneous profile as a porous SEI and the homogeneous profile as a dense SEI. Thus, our model predicts a transition from growth of dense to porous interphases as a function of film thickness. The universal origin of this effect is the reduction in growth rate with increasing film thickness. Low current densities favor heterogeneous growth, because the influence of fluctuations becomes larger as detailedly shown by Horstmann \text{et al.} \cite{Horstmann2018} and reflected by our stability criterion, Equation \ref{eq:instability_criterion}, confer SI for more detail.

\begin{figure}[tb]
    \centering
    \includegraphics[width=8.4cm]{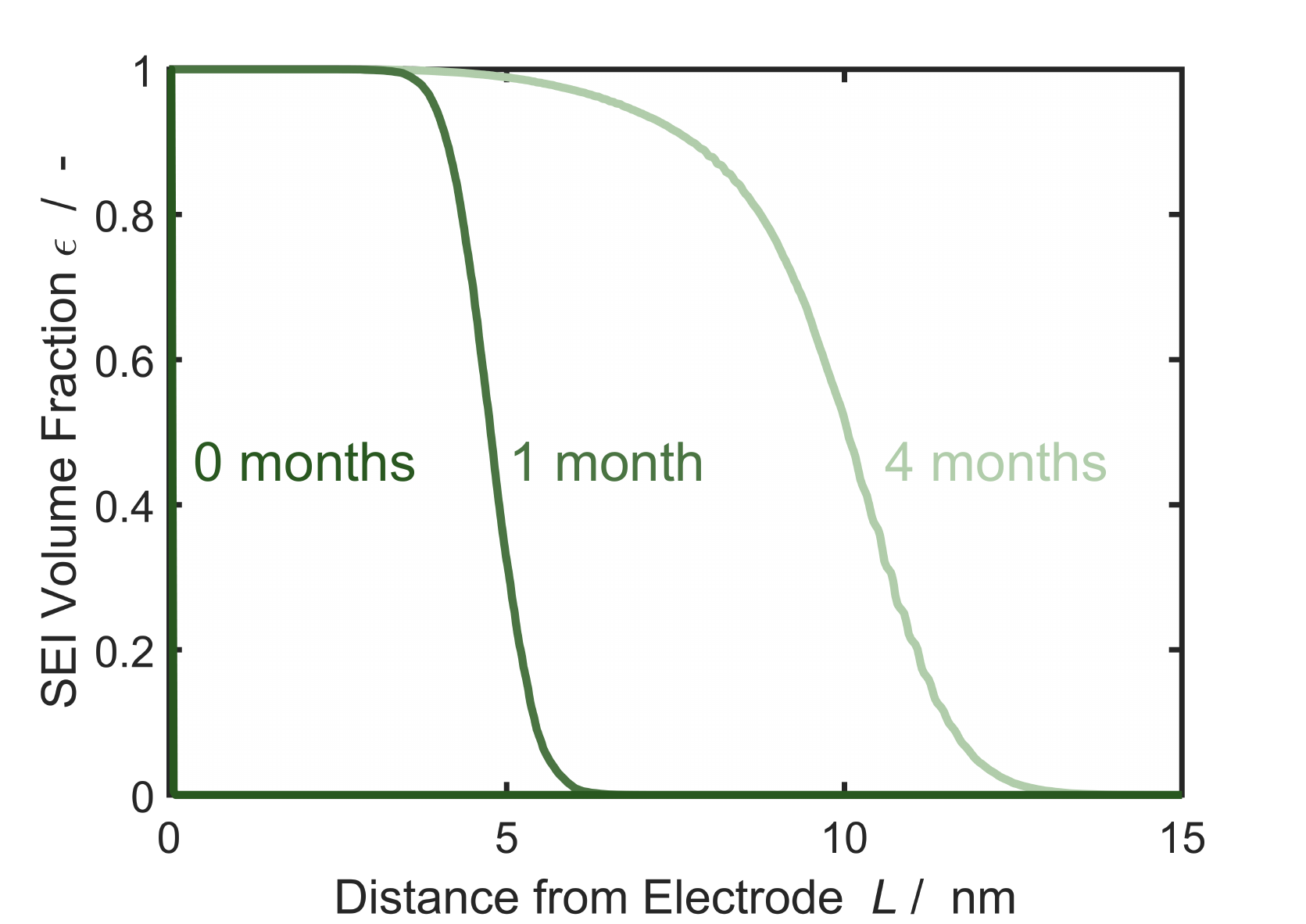}
    \caption{Evolution of the SEI volume fraction $\varepsilon_\SEI$ perpendicular to the electrode over time according to the morphology profiles of Figure \ref{fig:Morphology_Evolution}, averaged over 1000 simulations. Color gradient indicates increasing storage time from 0 (darkest green) to 4 months (lightest green).}
    \label{fig:Porosity_Evolution}
\end{figure}

To further characterize the SEI and compare it to existing theoretical studies, we investigate the SEI volume fraction $\varepsilon_\SEI$ in the following.
\begin{equation}
\varepsilon_\SEI(L)=\frac{1}{A}_x\int_0^{A_x} \xi(L) \text{d}x \qquad \xi(L) = 
\begin{cases}
1, \text{SEI at L} \\
0, \text{else}
\end{cases}
\end{equation}
Note that SEI volume fraction and porosity are closely connected, $\varepsilon=1-\varepsilon_\SEI$. Figure \ref{fig:Porosity_Evolution} shows the SEI volume fraction corresponding to the storage times shown in Figure \ref{fig:Morphology_Evolution}. We clearly see that the initially sharp interface between SEI and electrolyte smears out over time and after 4 months, we see a transition from $\varepsilon_\SEI=1$ to  $\varepsilon_\SEI=0$ along a thickness of approximately $\SI{8}{\nano\meter}$.

We interpret this morphological difference as emergence of a dual-layer SEI structure. In the following, we thus subdivide the SEI into a dense inner part of thickness $L_\text{in}$ without pores $\varepsilon_\SEI(L_\text{in})=1$ and a porous outer part $L_\text{out}=\max(L)-L_\text{in}$. The thickness $L_\text{in}$ is defined such that $\varepsilon_\SEI(L_\text{in}+a)<1$. 
\begin{figure}[tb]
    \centering
    \includegraphics[width=8.4cm]{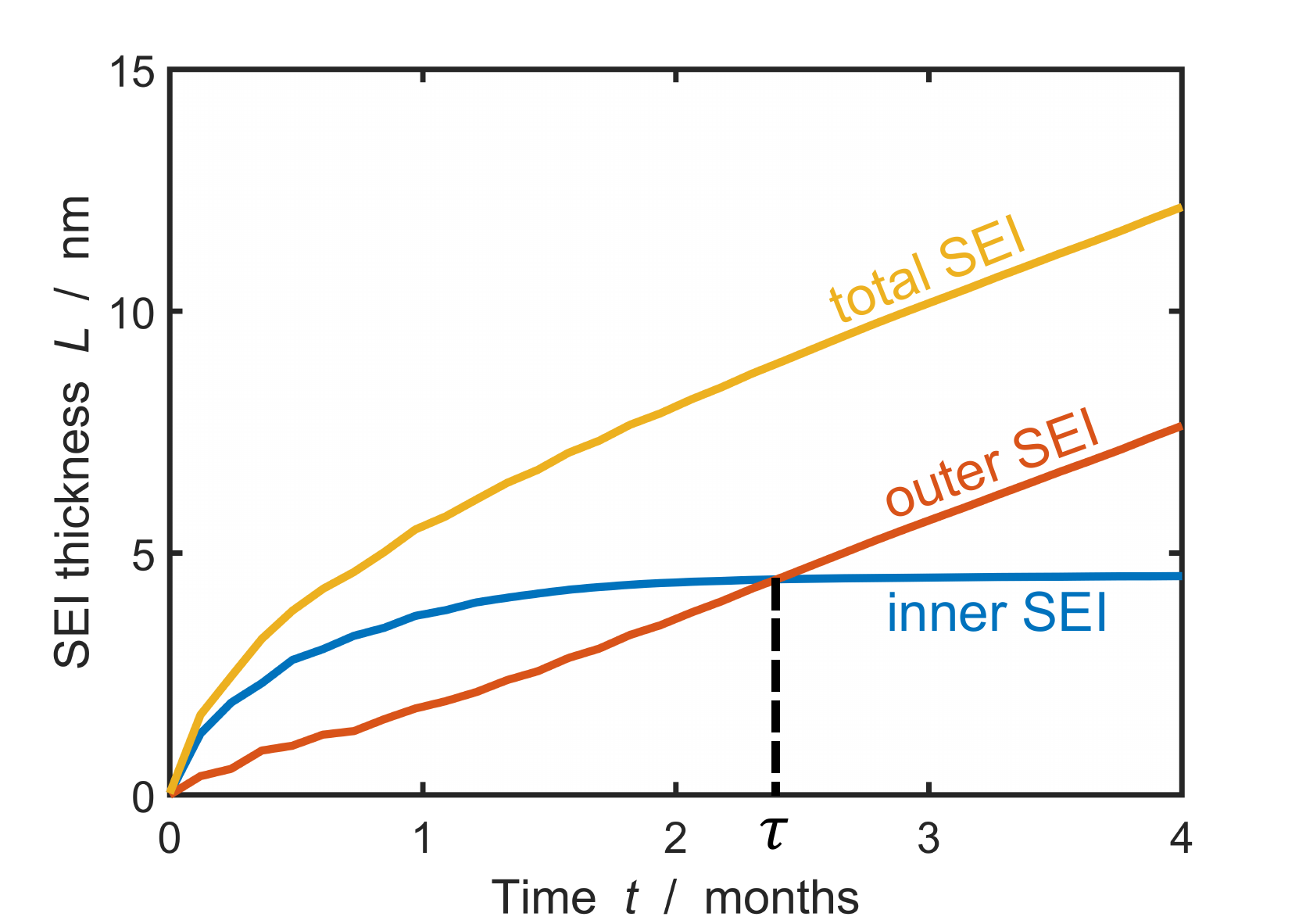}
    \caption{Growth of the inner (blue), outer (red) and total (yellow) SEI during storage for one year at 50\% state of charge. The results were averaged over 1000 simulations.}
    \label{fig:Dual_Layer}
\end{figure}

Figure \ref{fig:Dual_Layer} depicts the evolution of the inner and outer SEI at 50\% SoC. While the inner SEI reaches a steady thickness of $\SI{5}{\nano\meter}$ after 2 months, the outer SEI grows uninhibited, \textit{i.e.}, approximately linear in time. As a result, we observe the deviation from the parabolic growth law shown in Figure \ref{fig:sqrt_t_growth} at long times. 

This result clearly shows the onset of porous layer growth, but the long-term simulations have to be interpreted with care. However, our model cannot faithfully capture the long-term evolution after this transition for two reasons. First, we assume a constant lithium atom concentration $\tilde{c}$ at the SEI-electrolyte interphase. This assumption is reasonable for homogeneous SEI profiles and fast surface diffusion, but falls short with increasing film heterogeneity. Replacing the integral constraint Equation \ref{eq:Integral_Closing} with a fully resolved three-dimensional diffusion-reaction equilibrium would increase the model accuracy, but result in higher computational costs. Second, the heterogeneous film tends to arrange in columns and valleys getting stuck at unfavorable molecular positions in the SEI. At this point, we emphasize that this particular process stops the inner SEI growth in Figure \ref{fig:Dual_Layer}. This does not imply that the electrode is completely passivated and in reality we would expect the inner SEI to continue growing albeit at a significantly slower rate. It can be expected that neighboring columns can merge in real SEI and form truly porous structures. Such structures cannot be described with the thickness function $L(x)$ and would require a numerically more challenging tracking of the SEI surface.

The dual-layer SEI growth corresponds well to experimental findings. Harris and coworkers \cite{Lu2011,Harris2013,Lu2014} showed the dual-layer SEI structure using TOF-SIMS, isotope tracer experiments, and EIS. With cryo-TEM measurements, Cui and coworkers visualized a dense inner SEI on graphite, silicon and lithium \cite{Li2018,Huang2019,Huang2019b,Huang2020}.  
Also the SEI growth model of Single \textit{et al.} \cite{Single2016,Single2017} predicts the emergence of a dual-layer SEI.
However, this model relates the two SEI layers to two reduction reactions and two SEI components. Our current model, instead, derives it from a universal transition in growth mode for decreasing growth rates. As a result, Single \textit{et al.} predict a constant ratio between inner and outer SEI thickness instead of a constant inner SEI thickness.
An analysis of existing SEIs after distinct storage times \textit{e.g.} by transmission electron microscopy or neutron reflectometry could reveal the precise morphology evolution and thereby help to identify the origin of dual-layer growth.
 
\begin{figure}[tb]
    \centering
    \includegraphics[width=8.4cm]{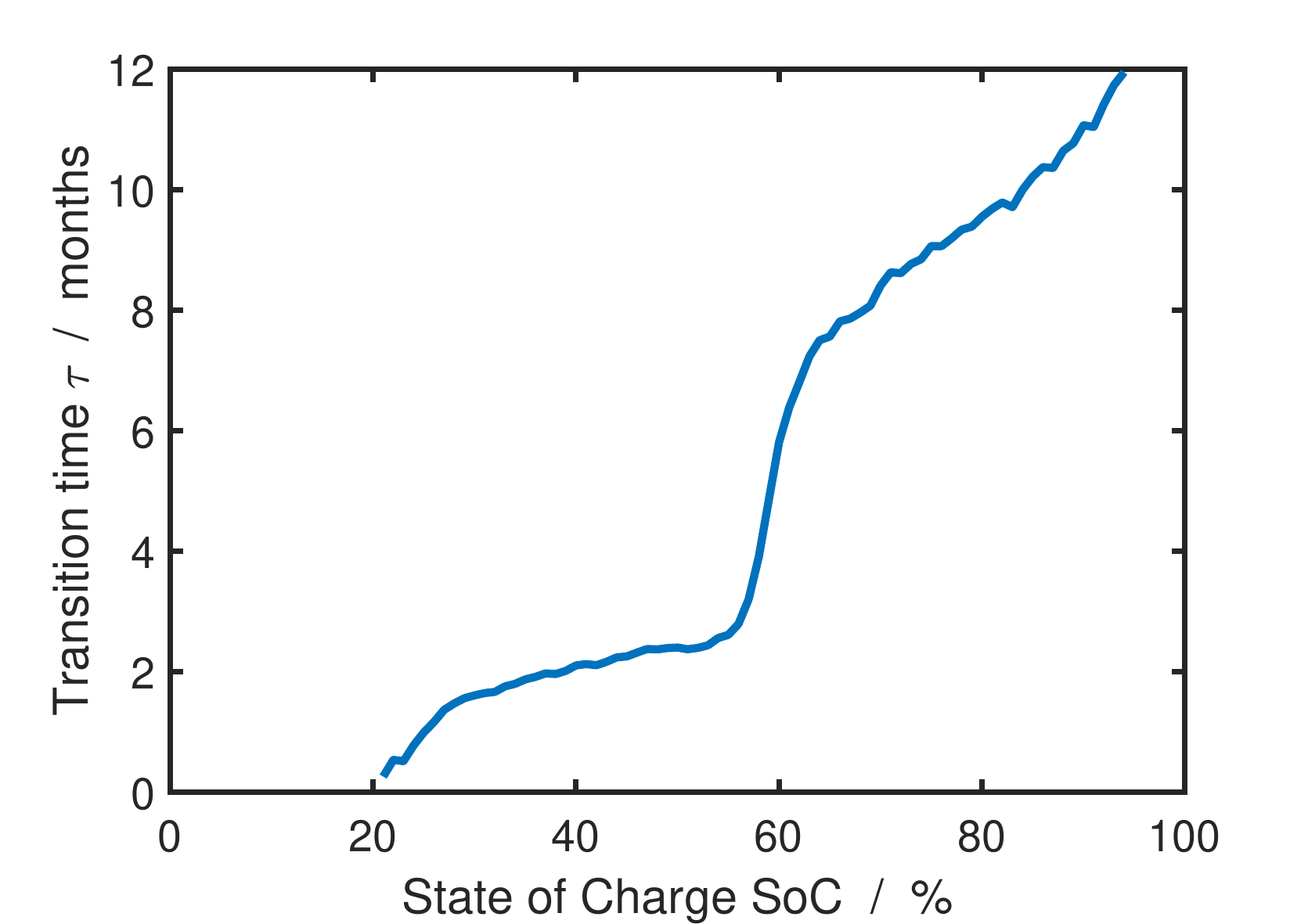}
    \caption{Transition time $\tau$ from dense to porous SEI growth defined as $L_\text{in}(\tau)=L_\text{out}(\tau)$ depending on the state of charge. The results were averaged over 1000 simulations.}
    \label{fig:Transition_Time}
\end{figure}

Finally, we analyze the influence of SoC on the emerging SEI profile during storage. To this aim, Figure \ref{fig:Transition_Time} shows the transition time $\tau$ from dense to porous growth, which we define as $L_\text{in}(\tau)=L_\text{out}(\tau)$ (see Figure \ref{fig:Dual_Layer}). 
We observe that the transition time increases with increasing state of charge. Accordingly, low SoCs result in a predominantly porous SEI, while high SoCs will lead to a dense SEI. This is a direct consequence of our model, which limits SEI growth by charge transport via electron diffusion. Electron concentration in the SEI at the electrode depends on the SoC \textit{via} the OCV-curve. Thus, electron transport through the SEI is faster at higher SOCs according to $N_\text{diff}\propto \exp{\left(-eU_0/k_\text{B}T\right)}$. As a consequence, lower SoCs cause slower electron transport, which in turn causes earlier onset of heterogeneous growth, according to our instability criterion, Equation \ref{eq:instability_criterion}. Our findings accord qualitatively well to the experiments of Harris and coworkers \cite{Lu2011,Harris2013,Lu2014}, who describe an increase in SEI porosity with increasing electrode voltage, \textit{i.e.} decreasing SoC. Note that the transition time $\tau$ increases with decreasing exchange current density of electrolyte reduction $r_0$, because this shifts the onset of porous growth as explained in SI-2. At this point, one could intervene in the experiment, e.g. by using additives, in order to delay the onset of porous growth.

\begin{figure}[tb]
    \centering
    \includegraphics[width=8.4cm]{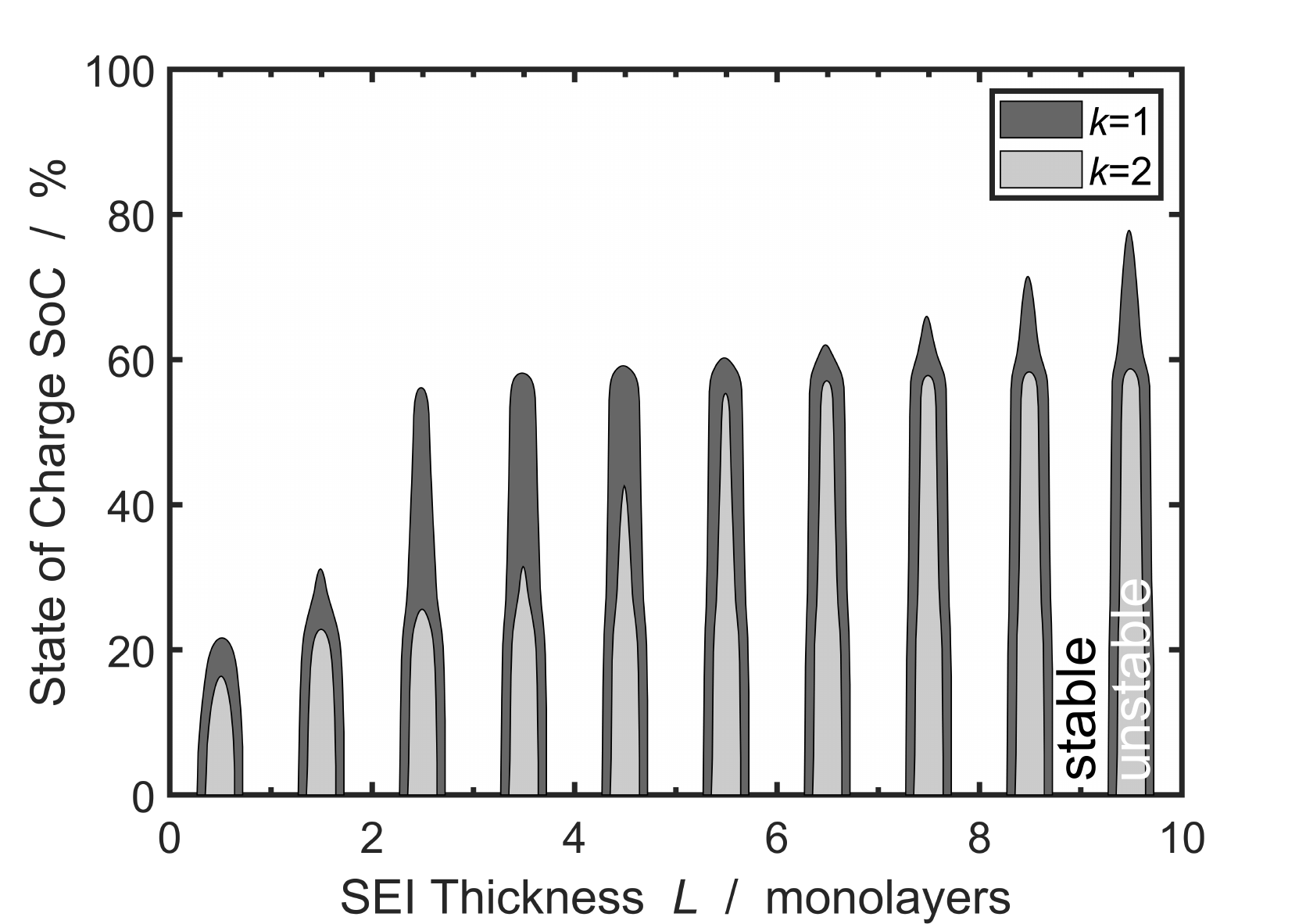}
    \caption{Stability analysis of SEI growth, according to Equation \ref{eq:Exponential_Growth_Rate}. Filled areas show conditions of unstable SEI growth $\tilde{s}>\partial \tilde{L}/\partial \tilde{t}$ depending on the SEI thickness $h$ and the state of charge for increasing wavenumbers $\tilde{k}$.}
    \label{fig:Stability_Analysis}
\end{figure}

We understand the influence of time and SoC on SEI morphology with infinitesimal stability analysis.
In Figure \ref{fig:Stability_Analysis}, we analyze the occurrence of metastable growth $\tilde{s}>\partial \tilde{L}_0/\partial \tilde{t}$ (see Equation \ref{eq:Exponential_Growth_Rate}) at different times, \textit{i.e.}, SEI thicknesses $L$ and SoCs. 
We observe three main trends in Figure \ref{fig:Stability_Analysis}. 
First, SEI growth is stable for integer monolayers. This effect results from the form of the Gibbs free energy, Equation \ref{eq:Free_Energy}, with Equilibria at complete molecular monolayers $\tilde{h}$ and a free energy barrier $E_1$ in between.
Second, SEI growth becomes more unstable over time, \textit{i.e.}, with increasing thickness $L$. This explains the increasingly porous SEI morphology, which we observed in Figure \ref{fig:Morphology_Evolution} and \ref{fig:Porosity_Evolution}.
Third, higher SoCs lead to more stable SEI growth. This accords well to our findings from Figure \ref{fig:Transition_Time}, which show a denser SEI for high SoCs.
These trends are analogous to the current dependent morphology evolution observed in $\ce{Li_2O_2}$ \cite{Horstmann2013}. Large leak currents, resulting either from a high SoC or a thin SEI, lead to dense SEI growth, while small leak currents cause porous SEI growth.

Overall, the stability analysis depicted in Figure \ref{fig:Stability_Analysis} comprises the main features of our model: SEI growth becomes increasingly unstable over time and with lower SoC. 
Thereby, our model shows that the experimentally observed dual-layer SEI structure not only results from the formation of different SEI compounds, but is also inherent to the growth process. In a more general view,our model can also be applied to analyse multi-layer formation in different aging phenomena, \textit{e.g.} oxidation of silicon, or patina formation on rocks or metals.

In conclusion, our model predicts a universal transition to dual-layer growth for passivating films. However, our predictions are rather qualitative as our model has some limitations. 
First, the model parametrization is difficult, because information about the long-term SEI morphology evolution is scarce. Novel TEM images of SEI layers after different battery storage times can reveal the structural evolution and aid in parametrizing the model. Moreover, these experiments could measure growth of the inner and outer layer separately and distinguish between the model predictions presented in this paper and by Single \textit{et al.} \cite{Single2016,Single2017}.
Second, the combination of effects on different length and time scales challenges our model, which is highly sensitive to molecular fluctuations as described by the stability analysis. Our reductionist model focuses on the SEI structure and thus only captures randomness emerging from structural disorder. Additional fluctuations, \textit{e.g.}, thermal, chemical, or electrolyte related may influence our predictions. Further information about these fluctuations as well as surface energies from ab-initio calculations would improve our model.

\section{Conclusion}

Summing up, we developed a general model to study the morphology evolution of passivating surface films. The model consists of a diffusion equation and a morphology-driven reaction rate \cite{Single2018,Horstmann2013}.
These films arise for example on crystalline silicon \cite{Nunomura2020}, minerals \cite{Garcia-Valles1998}, and metals \cite{Domenech-Carbo2019,Thomas2013,Macdonald1999}.

We use the model to study the morphology evolution of the solid-electrolyte interphase (SEI) on anode particles in lithium-ion batteries. We parameterize the state-of-charge and time dependence of SEI growth with capacity fade experiments \cite{Keil2016,Keil2017}. The resulting model captures the SEI's tendency to grow in a dual-layer structure with a dense inner and a porous outer layer and captures the principles of experimental findings on the voltage dependence of this process \cite{Lu2011,Lu2014,Harris2013}.

Our model focuses on the onset of porous layer growth. Modelling the long-term growth of porous structures would require a simulation of two-dimensional diffusion through the SEI. This approach would be interesting for future studies but comes at the cost of additional unknown parameters. Novel in-situ experiments could provide improved model parameterization and validation. 

\section*{Author Contributions}
Lars von Kolzenberg: Conceptualization, Methodology, Software, Validation, Formal analysis,  Investigation, Writing - Original Draft, Visualization

Martin Werres: Conceptualization, Methodology, Software, Validation, Formal analysis,  Investigation, Writing - Original Draft, Visualization

Jonas Tetzloff: Methodology, Software, Formal analysis,  Writing - Review \& Editing

Birger Horstmann: Conceptualization, Methodology, Writing - Review \& Editing,  Supervision, Funding acquisition, Project administration

  \section*{Acknowledgements}
    Lars von Kolzenberg gratefully acknowledges funding and support by the German Research Foundation (DFG) within the research training group SiMET under the project number 281041241/GRK2218. Martin Werres gratefully acknowledges financial support within the Lillint project (03XP0225A). The support of the bwHPC initiative through the use of the JUSTUS HPC facility at Ulm University is acknowledged. This work contributes to the research performed at CELEST (Center for Electrochemical Energy Storage Ulm-Karlsruhe).

\section*{Conflicts of interest}
There are no conflicts to declare.





\bibliography{literatur_abbr} 

\end{document}